\documentclass[prb,twocolumn,showpacs,preprintnumbers,amsmath,amssymb]{revtex4}
\usepackage{graphicx}
\usepackage{amsmath}
\usepackage{color}

\begin{document}

\title{Inverse freezing in the Ghatak-Sherrington model with a random field.}

\author{C. V. Morais and M. J. Lazo  }

\affiliation{Programa de P\'os-Gradua\c{c}\~{a}o em F\'{\i}sica - Instituto de Matem\'atica, Estat\'{\i}stica e F\'{\i}sica, 
Universidade Federal do Rio Grande, 
96.201-900, Rio Grande, RS, Brazil }

\author{F. M. Zimmer}
\affiliation{Departamento de Fisica, Universidade Federal de Santa Maria,
97105-900 Santa Maria, RS, Brazil}

\author{S. G. Magalhaes}
\affiliation{Instituto de Fisica, Universidade Federal Fluminense, 24210-346
Niter\'oi, RJ, Brazil}

\date{\today}

\begin{abstract}

The present work studies the Ghatak-Sherrington (GS) model in the presence of a magnetic random field (RF). Previous results obtained from GS model without RF suggest that disorder and frustration are the key ingredients to produce spontaneous inverse freezing (IF). However, in this model, the effects of disorder and frustration always appear combined. In that sense, the introduction of RF allows us to study the IF under the effects of a disorder which is not a source of frustration. The problem is solved within the one step replica symmetry approximation. The results show that the first order transition between the spin glass and the paramagnetic phases, which is related to the IF for a certain range of crystal field $D$, is gradually suppressed when the RF is increased. 

\end{abstract}

\maketitle

\section{Introduction}

In recent years, there is considerable interest in an unusual class of phase transitions known as Inverse Transitions (IT). In these counterintuitive transitions, the ordered phase has more entropy than the disordered one. It is important to remark that such phase transitions have been observed experimentally in physical systems as, for example, magnetic thin films \cite{8} and high-Tc superconductors \cite{7}. From the theoretical point of view, the understanding of the puzzling mechanisms behind such transitions is a challenging task. In this sense, some magnetic models can be quite useful (see, for instance, Refs. \onlinecite{9,10,11,12,13,14,15,16,17}). Indeed, in some of these mentioned references, it is suggested that disorder and frustration are key ingredients to produce spontaneously IT and, in particular, Inverse Freezing (IF). In this kind of IT, the spin glass (SG) phase is the ordered one. Nevertheless, in most of these spin models, disorder appears only as a random bond with no other forms of disorder being considered. 

A new form of disorder that could be added to these models is a random field. Albeit the addition of a random field in these spin models is a new simple term, it can introduce a complex interplay of two highly non-trivial manifestations of disordered magnetism, i. e., spin glass and random fields \cite{21,22}. As an example of this complex interplay, recent studies of the Sherrington-Kirkpatrick (SK) model \cite{26} with a random field indicate that the SG phase is deeply affected by the presence of such field. In particular, not only the nature of the SG boundary phase can be affected \cite{27,28,29} but also its location can be changed \cite{30,31}. On the other hand, it should be noticed that to increase the disorder in the problem by adding a random field does not mean necessarily to increase frustration. In fact, as shown by the Random Field Ising model (RFIM) in the mean field approximation, the presence of a random field is an additional form of disorder, nonetheless, it is not an additional source of frustration \cite{Pyte,Aharony}.

An example of a simple disordered magnetic model presenting spontaneously IF in the mean field approximation is the Ghatak-Sherrington (GS) model \cite{18}. The precise meaning of spontaneous existence of IF is that in this model it is not used any artificial procedure to increase the entropy of the SG phase \cite{9}. From the previously mentioned results for the SK model and  the RFIM, one could ask what would happen with the spontaneous IF found in the GS model when a random field is applied? Is the reentrance in the first order boundary between the SG and paramagnetic phases affected by the presence of the random field? The answer to  these questions can not only be useful to clarify the role of disorder as an ingredient to produce (or destroy) an IF, but also to distinguish the effects coming from disorder, from those ones coming from frustration. The reason is that, at mean field level, the effects of a local disorder coming from random fields can be distinct from those ones coming from random bonds. For this last case, the effects of disorder and frustration can appear combined, as in the GS model.
 
In the present work we study the GS model in the presence of a longitudinal random field. In this model, the spins can assume values $S=\pm 1, 0$, $D$ is a crystal lattice field and the spin-spin coupling $J_{ij}$ and the random field $h_i$ follow Gaussian distributions. In the case of a random field following a Gaussian distribution, it is necessary to use the replica symmetry breaking in order to obtain the SG solution \cite{31}. The free energy is obtained and consequently, the SG order parameters. The disorder is treated using the replica method in which the SG solutions are searched within the one-step replica symmetry breaking (1S-RSB) approximation \cite{Parisi}. The location of the first order SG/PM boundary phase, in the present work, is done by using the same criteria proposed in Ref. \onlinecite{salinas}. In this case, the chosen SG solution in the transition is that one which meets continuously the SG solution at small $D$. This solution corresponds to the lower SG free energy.  It is important to point out that the GS model, besides disorder and frustration, has the ingredients, as proposed by Schupper and Shnerb \cite{9}, to present spontaneous IT, i.e., an interacting term favoring a magnetic state and a mechanism favoring non-magnetic states in the sites.

It should be remarked that in the GS model there is a new order parameter $p$, which is related to the diagonal replica components (the spin self-interaction $p=\langle\langle S^{\alpha }S^{\alpha }\rangle\rangle_{J,h}$). In fact, this order parameter is related to the average number $n_0=1-p$ of non-magnetic states in the sites which also reflects the capability of the sites to interact or not \cite{Castillo}. For example, the limits $n_0=0$ and $n_0=1$ can be related to the fully interacting ($p=1$) and the fully noninteracting ($p=0$) regimes, respectively. The behavior of $n_0$ can be important to clarify the mechanism behind the existence of the IF in the present approach. 

The paper is structured as follows: in section II, the free energy within 1S-RSB scheme was found. In section III, a detailed discussion of phase diagrams is presented. The last section is reserved to the conclusions.

\section{General Formulation}

The model is given by the Hamiltonian below
\begin{equation}
H = - \sum_{(i,j)} J_{ij}  S_{i}  S_{j} + D \sum_{i = 1 }^{N} S_{i}^2 - \sum_{i} h_{i}  S_{i},
\label{eq1}
\end{equation}
where the spin can assume values $S=\pm 1,0$, the summation $(i,j)$ is over any two interacting sites and $D$ is the crystal field. The spin-spin coupling $J_{ij}$ and the magnetic fields $h_i$ are random variables following independent Gaussian distributions 
\begin{equation}
P(J_{ij})=\left[\frac{N}{2\pi J^{2}}\right]^{1/2} \exp\left[-\frac{N}{2J^{2}}J_{ij}^{2}\right]
\label{eq2a}
\end{equation}
\begin{equation}
P(h_i)=\left[\frac{1}{2\pi \Delta^{2}}\right]^{1/2} \exp\left[-\frac{1}{2\Delta^{2}} h_i^{2} \right]\\.
\label{eq2}
\end{equation}
Herein the procedure introduced in the Refs. \onlinecite{30,31} is closely followed. Therefore, the free energy per spin can be obtained using the replica method as 
\begin{equation}
-\beta f=\lim_{N\rightarrow\infty}\lim_{n\rightarrow 0}\frac{1}{Nn}\left(<<Z^{n}>>_{J,h}-1\right)\\,
\label{eq4}
\end{equation}
where the average $<<f (\left\{J_{ij}\right\},\left\{h_i\right\})>>_{J,h}$ is given by
\begin{equation}
\begin{split}
&<< f (\left\{J_{ij}\right\},\left\{h_i\right\})>>_{J,h}=\int \prod_{(i,j)}\left[dJ_{ij}P(J_{ij})\right]\\ &\times
\prod_i\left[dh_{i}P(h_i)\right] f(\left\{J_{ij}\right\},\left\{h_i\right\}),
\end{split}
\label{eq3}
\end{equation}
with $J_{ij}$ and $h_i$ following the Gaussian distributions defined in Eq. \ref{eq1}.

From Eq. \ref{eq4}, after averaging $Z^{n}$ over $J_{ij}$ and $h_{i}$, one gets by using the Hubbard-Stratonovich transformation the following relation:
\begin{equation}
\label{eq5}
 \begin{split}
  - \beta f 
&= \lim_{n\rightarrow 0} \frac{1}{n}  \left\lbrace 
- \frac{(\beta J)^2}{2} \sum_{(\alpha\beta)} q^{2}_{\alpha\beta} \right.   
\\&- \left. 
\frac{(\beta J)^2}{4} \sum_{\alpha} q^{2}_{\alpha\alpha}+ \ln \mbox{Tr} e^{L}  
\right\rbrace,
 \end{split}
\end{equation}
where $L$ is the effective Hamiltonian
\begin{equation}
\label{eq7}
\begin{split}
L &=   (\beta J)^{2} \sum_{(\alpha\beta)} q_{\alpha\beta} S_{\alpha\beta} + (\beta \Delta)^2 \sum_{(\alpha\beta)} S_{\alpha}S_{\beta} \\&+ \left( \frac{(\beta \Delta)^2}{2} - \beta D \right) \sum_{\alpha} (S_{\alpha})^2 + \frac{(\beta J)^2}{2} \sum_{\alpha}q_{\alpha\alpha}(S_{\alpha})^2,
\end{split}
\end{equation}
and the diagonal $q_{\alpha\alpha}$ and non diagonal $q_{\alpha\beta}$ elements of the replica order parameter matrix are given by
\begin{equation}
q_{\alpha\alpha}=<S_{\alpha}^2>,\;\;\; q_{\alpha\beta}=<S_{\alpha} S_{\beta}>,
\label{eqnova2}
\end{equation}
respectively. The symbol $<...>$ means the thermodynamical average over the effective Hamiltonian $L$ given in Eq. \ref{eq7}. 

The Parisi's 1S-RSB\cite{Parisi} is used in Eq. \ref{eq5}. Therefore, the replica matrix ${q}$ is parametrized as shown below \cite{Parisi}
\begin{equation}
q_{\alpha\beta}=
q_{1} \mbox{ if } I(\alpha/x)=I(\beta/x)
\label{eq8.1}
\end{equation}
\begin{equation}
q_{\alpha\beta}=q_{0} \mbox{ if } I(\alpha/x)\neq I(\beta/x)
\label{eq8.2}
\end{equation}
where $I(x)$ gives the smallest integer which is greater than or equal to $x$.  While
\begin{equation}
q_{\alpha\alpha}=p.
\label{eqnova4}
\end{equation}
The parametrization introduced in Eqs. \ref{eq8.1}-\ref{eqnova4} is used in the sum over the replica labels in Eqs \ref{eq5}-\ref{eq7}.

The final step in the 1S-RSB is a linearisation in Eq. \ref{eq7} which introduces new auxiliary fields $z$ and $v$. Finally, the free energy is given by
\begin{equation} 
\begin{split}
\beta f(q_0,q_1,p,x)&= \frac{(\beta J)^{2}}{4}[(x-1)q_{1}^{2} - x q_{0}^{2} +p^{2}] 
\\& - \frac{1}{x} \int Dz  \ln \left[ \int Dv \left[K(z,v)\right]^{x} \right],
\label{eq1aaasd}
\end{split}
\end{equation} 
where
\begin{equation}
K(z,v)= 1 + 2 e^{\gamma}  \cosh h(z,v)
\end{equation}
with $\int D w=\int\frac{d w}{\sqrt{2\pi}}e^{-w^{2}/2}$ $(w=z,v)$, 
\begin{equation}
\gamma = \frac{(\beta J)^2}{2} (p - q_1) - \beta D,
\end{equation}
and $h(z,v)$ is
\begin{equation} 
\begin{split}
h(z,v)&=\beta J[ \sqrt{  q_0 + \Delta^2/J^2}z + \sqrt{ q_1-q_0}v ].
\end{split}
\label{e30sgs}
\end{equation}

The equations for the order parameters  $q_{0}$, $q_{1}$, $p$ and  the block size parameter $x$ are obtained from Eq. \ref{eq1aaasd} by using the saddle point conditions. Other thermodynamic quantities can also be derived from Eq. \ref{eq1aaasd}, as for instance, the entropy $s=-\partial f/\partial T$.

\section{Results}

In this section, the numerical results for the order parameters are displayed 
\begin{figure}[ht!]
\includegraphics[width=5.6cm,angle=-90]{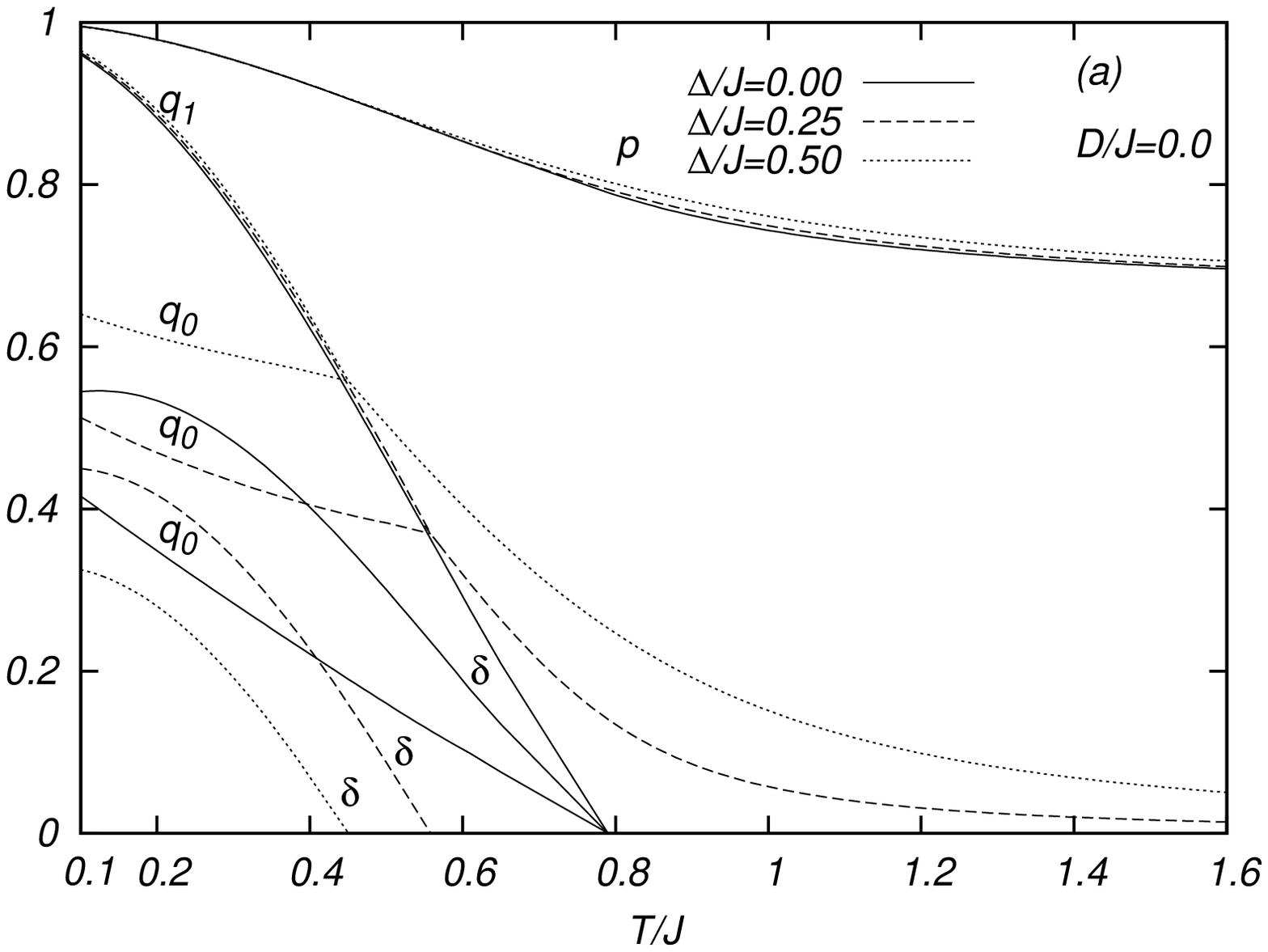} 
\includegraphics[width=5.6cm,angle=-90]{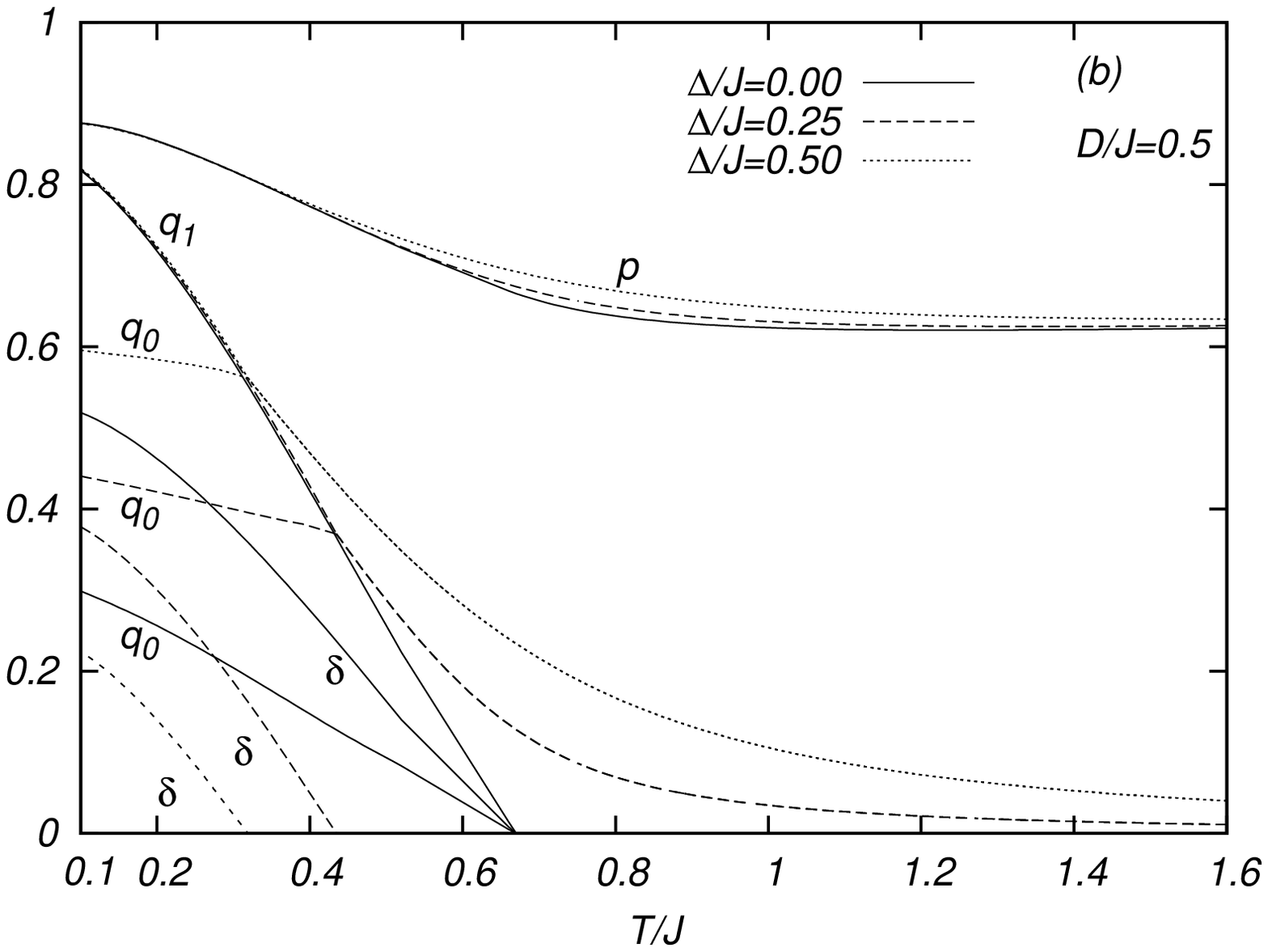}
\caption{Order parameters as a function of temperature for $D/J=0$ (upper panel),   $D/J=0.5$ (lower panel) and several values of $\Delta$.}
\label{fig1}
\end{figure}
with different set of values of the width $\Delta$ (see Eq. \ref{eq2}) and the crystal field $D$ that are given in units of $J$. These results allow us to build several phase diagrams in which the SG phase is characterized by the RSB order parameter $\delta\equiv q_1-q_0>0$. For $\Delta>0$, the paramagnetic (PM) phase is given by $\delta=0$ but $q_{1}=q_{0}=q\neq 0$. The behavior of entropy and $n_0$ \cite{Castillo} are also presented.
\begin{figure}[ht]
\includegraphics[width=6cm,angle=-90]{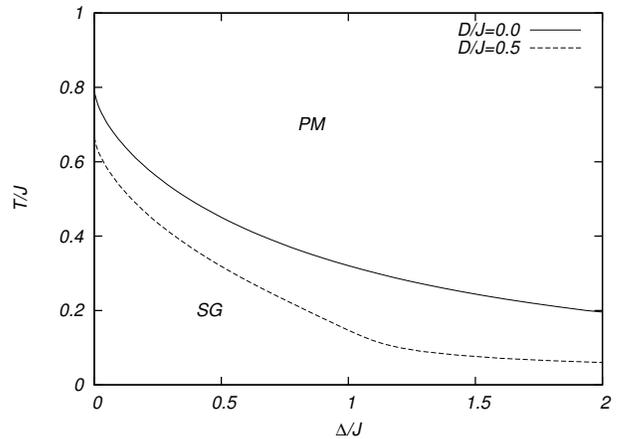} 
\caption{Phase diagrams of temperature versus $\Delta/J$ for $D/J=0.0$ (full line) and  $D/J=0.5$ (dashed line).}
\label{fig2}
\end{figure}
\begin{figure*}[ht]
\includegraphics[width=6cm,angle=-90]{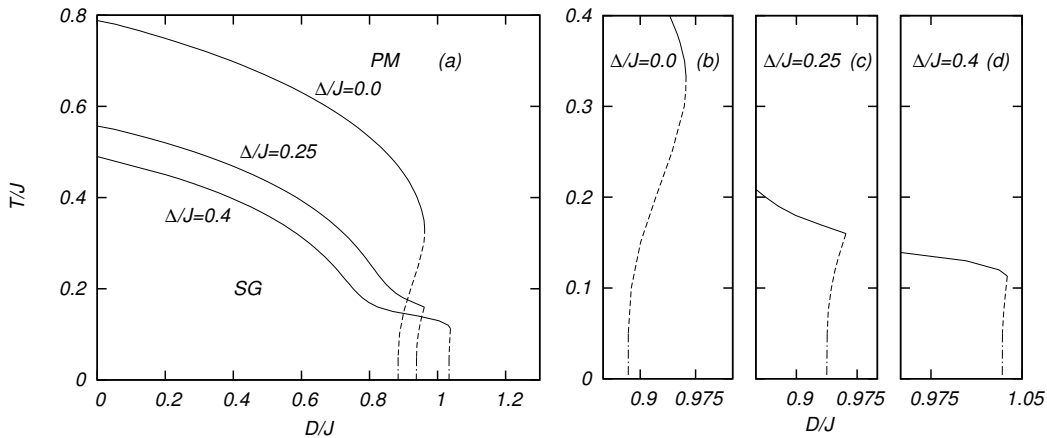}
\caption{Phase diagrams of temperature versus $D/J$ for several values of $\Delta$. Panels (b), (c) and (d) show details of the first order transitions. {The dashed-dotted lines at low $T$ are extrapolations to the $T=0$}. 
}
\label{fig3}
\end{figure*}
\begin{figure}[ht!]
\includegraphics[width=6cm,angle=-90]{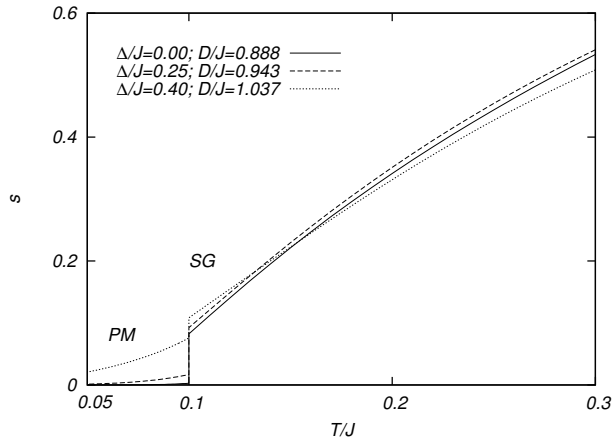}
\caption{Entropy as a function of temperature (region where occurs the reentrance) for three values of $\Delta$ and $D$.   
The first order temperature is equal to $T_{1f}=0.1J$ for all curves.
}
\label{fig4}
\end{figure}
Figure \ref{fig1} exhibits the behavior of order parameters for different configurations of $\Delta$ and $D$. In particular, the values chosen for $D$ corresponds to a range in which a PM/SG second order phase transition appears. For $\Delta=0$, it is obtained, as well known, that the temperature where $\delta>0$ (the freezing temperature $T_f$) is decreased when $D$ increases \cite{18}. The results also indicate that $p$ depends on the temperature, with every site becoming magnetic as $T\rightarrow 0$ for $D=0$ ($p\rightarrow 1$). Most important, for $D>0$, $p<1$ as $T\rightarrow 0$, which indicates that $D$ affects directly the average number of non-magnetic sites $n_0$. When $D=0$, the results are similar to those ones found recently for the SK model with a random field (see Ref. \cite{31}). For $\Delta=0$, the PM/SG phase transition is trivially induced. However, for $\Delta \neq 0$, the random field plays a role similar to a uniform magnetic field suppressing the SG solution in the replica symmetry (RS) scheme. In that case, the PM/SG phase transition can only be found in the 1S-RSB scheme ($\delta>0$). One important consequence is that $T_f$ decreases as $\Delta$ enhances. However, the spin self-interaction $p$ remains dependent on temperature but much less affected. Particularly, at very low temperatures. Finally, when the effects of finite $D$ and $\Delta$ are combined, some of the previous effects are preserved but others are more pronounced. For instance, the random field still acts as a uniform magnetic field suppressing the SG solution. Despite that, $T_f$ decreases faster and $p$ is more affected with the increase of $\Delta$ for intermediated temperatures than the case $D=0$.

The behavior of $T_f$ discussed above is better analyzed in the phase diagrams of Fig. \ref{fig2}. For instance, $T_f$ is gradually decreased by $\Delta$. However, the SG phase is always found at lower temperatures (see full line in Fig. \ref{fig2}). For $D/J=0.5$, $T_f$ appears at even lower temperatures. However, the results suggest that the SG phase can still be found even for very large values of $\Delta$. 

Fig. \ref{fig3} shows the results for larger values of $D$. In that case, the continuous PM/SG phase transition is decreased until a tricritical point as $D$ enhances.  After the tricritical point, the phase transition becomes first order where there are more than one PM and SG solutions.
The first order boundary is then located by comparing the free energy of the stable PM solution to the 1S-RSB free energy of the SG solution that presents lower free energy\cite{salinas}. Particularly, the SG/PM first order transition can be reentrant. This reentrance represents an IF, in which the usual entropic relation between the ordered and disordered phases is changed (see  Fig. \ref{fig4}). This result for $\Delta/J=0.0$ was already discussed in Ref. \onlinecite{9}. On the other hand, the presence of the random field affects significantly the first order phase transition line. The tricritical point is displaced to lower temperatures and higher $D$ as $\Delta$ increases. For $\Delta$ large enough, the tricritical point is not found in our range of temperatures as well as the SG/PM reentrance. However, for smaller values of $\Delta$, the reentrance is still observed for a certain range of $D$ (see Figs. \ref{fig3}(b)-(d)). Nevertheless, the reentrance is gradually diminished by $\Delta$. It suggests that although the disorder introduced by the random field is relevant to the location of critical lines, it is not able to enhance the IF region. On the contrary, the reentrant transition tends to be suppressed by the presence of a random field.

Fig. \ref{fig4} presents the entropy $s$ as a function of $T/J$ for three values of $\Delta/J$: $0.0$, $0.25$ and $0.40$. 
The values of $D/J$ are chosen in an order that they allow us to follow the entropy difference between the SG and PM phases at the same first order boundary temperature $T_{1f}=0.1J$. For $\Delta=0$, the entropy of the PM phase is found below the SG one at the first order transition. As already presented in Ref. \onlinecite{9}, this result confirms the existence of IF in the GS model. When there is a random field, the entropy at low $T/J$ increases especially the PM phase. As a consequence, the entropy difference of the SG and PM phases at the transition is decreased. This result confirms that the reentrance seems to be weakened by the random field.

\begin{figure*}[ht]
\includegraphics[width=6.2cm,angle=-90]{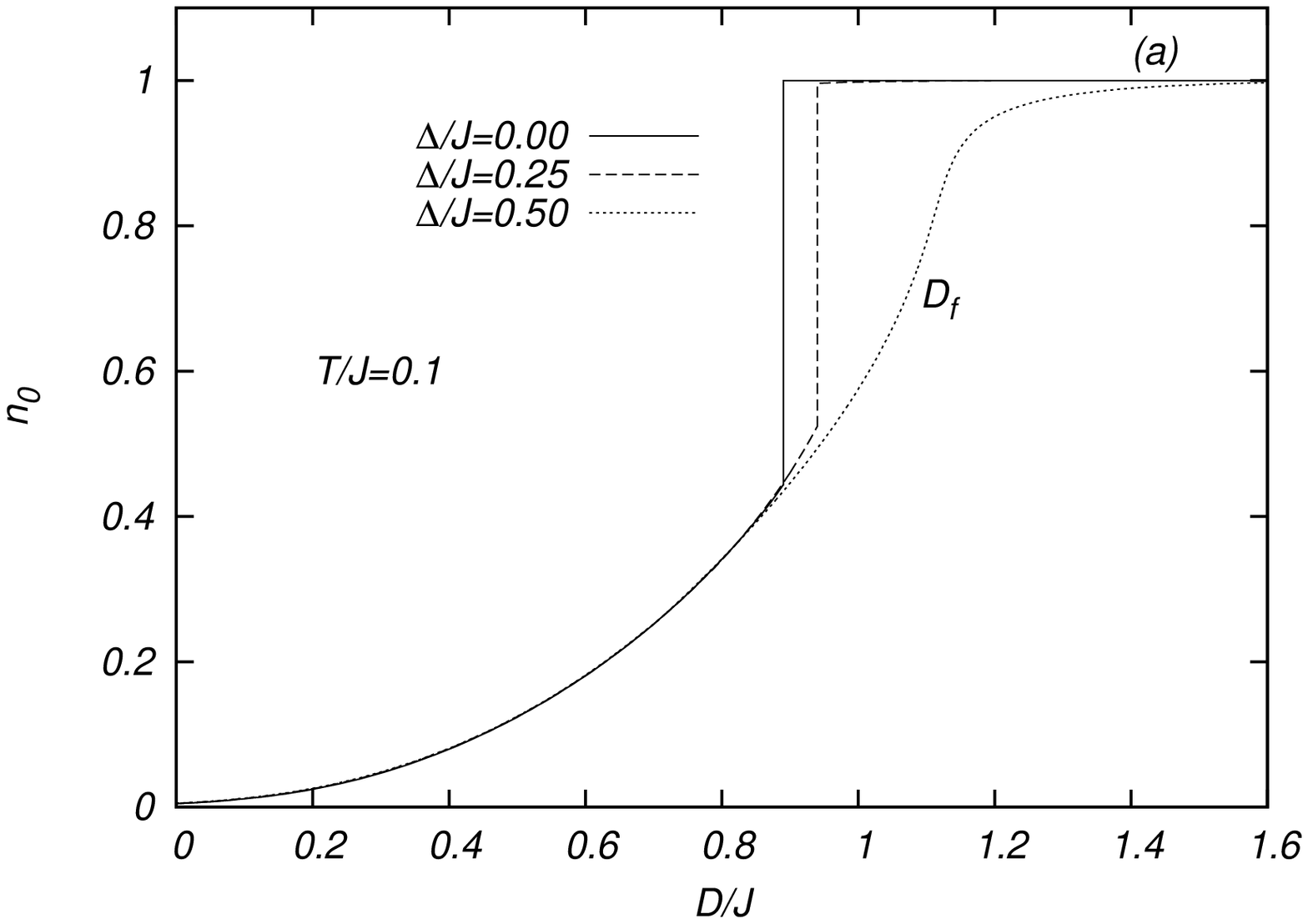}
\raggedright
\includegraphics[width=6.2cm,angle=-90]{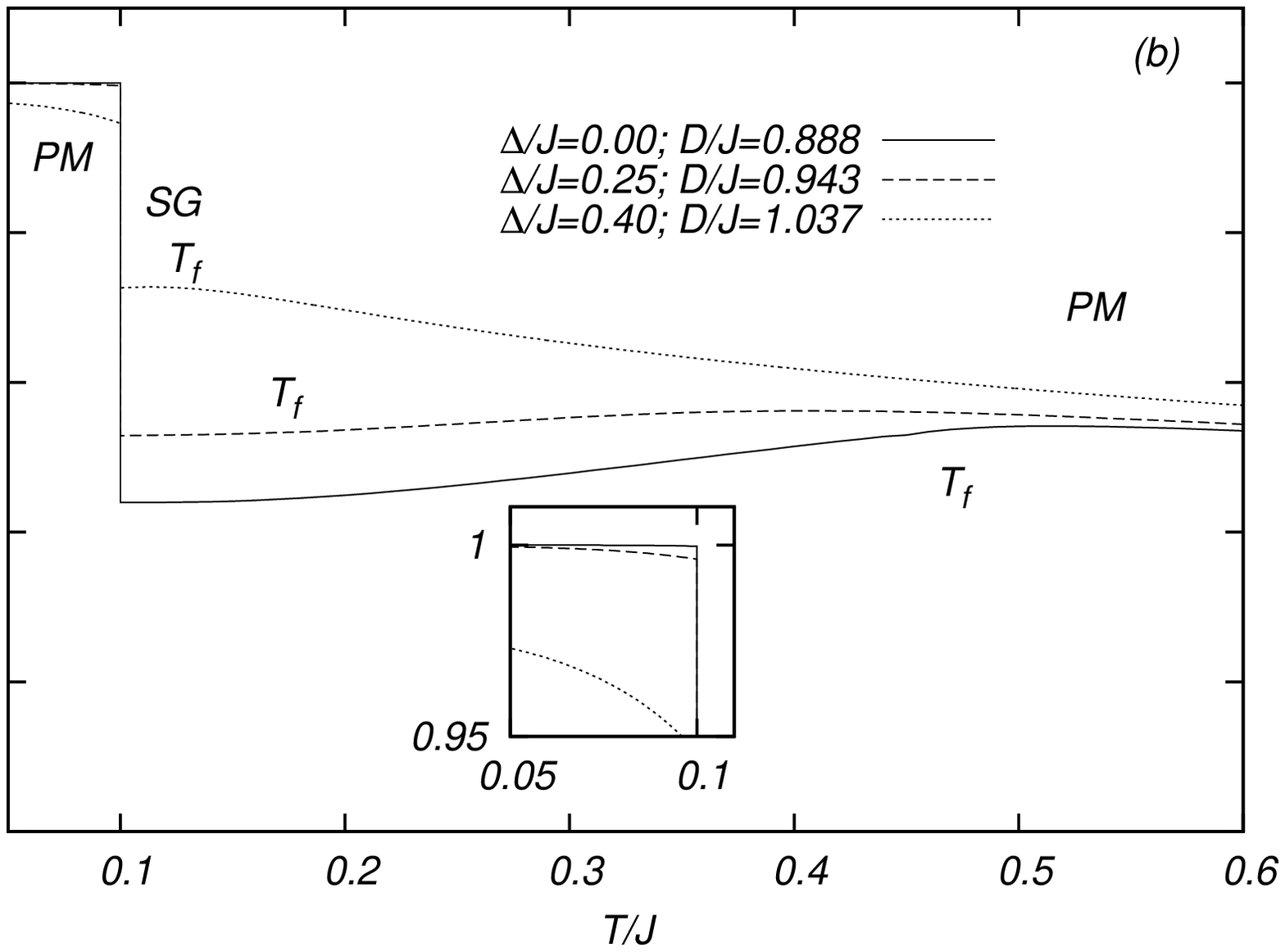}
\caption{Panel (a) shows the average number of sites with nonmagnetic states $n_0$ versus $D/J$ for $T/J=0.1$ and several values of $\Delta/J$. Panel (b) shows $n_0$ versus $T/J$ for three values of $D/J$ and $\Delta/J$. The values of $D/J$ in panel (b) are chosen in order to present the first order temperature equal to $T_{1f}=0.1J$. $T_f$ represents the freezing temperature.
}
\label{fig31}
\end{figure*}

Fig. \ref{fig31} exhibits $n_0$ which can be useful to understand the behavior of phase diagrams $T/J$ versus $D/J$ shown in Fig. \ref{fig3}. For the isotherm $T/J=0.1$ with $\Delta=0$ and $0.25$  shown in Fig. \ref{fig31}(a), $n_0$ increases continuously with $D$ until the PM/SG first order phase transition where it tends discontinuously to the higher $n_0$ limit (see full line in Fig. \ref{fig31}(a)). The behavior of $n_{0}$ in the SG is not affected by the random field. However, for $\Delta=0.25$, the first order phase transition is displaced for a larger value of $D$ as shown in Figs. \ref{fig3}(b)-(c). For $\Delta=0.5$, the tricritical point is located below the isotherm $T/J=0.1$. Therefore, the PM/SG phase transition becomes a second order one, with $n_0$ increasing continuously. However, $n_0$ in the PM phase is decreased as compared to the previous cases. This behavior could explain the presence of the SG order at a large range of $D$. To put in another way, $D$ favors the nonmagnetic states $(n_0 \rightarrow 1)$, but the random field enforces the magnetic ones. Therefore, the random field can diminish the effects caused by D.

Fig. \ref{fig31}(b) shows that $n_0$ also depends on the temperature. For $\Delta/J=0$, $n_0$ in the PM phase increases as the temperature diminishes until the $T_f$, where the SG order appears. However, the PM phase can be found again at lower temperatures. Most important, the behavior of $n_0$ in the PM phase at low and high $T$ are different. At low $T$, the $n_0$ of the PM phase presents high level of noninteracting states as compared to its counterpart at high $T$. As a consequence, the reentrance can appear. In other words, the nontrivial frustration and the presence of non interacting states can introduce an IF transition. On other hand, the disorder caused by the random field alters the $n_0$ distribution in such way that inverse freezing is weakened. It tends to decrease the difference between the $n_0$ of the PM and SG phases at the transition. In addition, the $n_0$ of PM phase at low temperature is decreased by $\Delta$, which results in the increase of the entropy of the PM phase at low $T/J$, as shown at Fig. \ref{fig4}.

\section{Conclusions}

In the present work, the Ghatak-Sherrington (GS) spin glass model has been studied with the addition of a random longitudinal 
magnetic field $h_{i}$ and a crystal field $D$. This study was carried out within the 1S-RSB scheme using a mean field 
approximation. It is known that spontaneous Inverse Freezing (IF) can appear in this model depending on the value given for 
$D$ \cite{18}. The main purpose of the present work is to investigate the effects of the random field following a Gaussian 
distribution in the IF. Since, it does not introduce any additional frustration in the problem (at least, at mean field level 
\cite{13}) its presence can be helpful to elucidate what is the role of disorder as an ingredient to produce (or suppress) 
the IF as distinct of frustration.

Our results show that the random field depresses the freezing temperature $T_f$. Moreover and most importantly for the 
purposes of the present work, the random field decreases the reentrance in the first order phase transition line PM/SG. The 
mechanism responsible for such effect is concerned with the average number of sites with nonmagnetic states in the SG and PM 
phases. It is important to remark that in the absence of a random field, the PM phase at low temperature is characterized by 
higher number of sites with nonmagnetic states as compared with the PM phase at higher temperature. By contrast, the random 
field tends to diminish the number of sites with nonmagnetic states of the PM phase at lower temperature. In that way, the SG 
phase is again energetically favored at lower temperature.  

Therefore, one can conclude that the disorder given by the presence of a random field does not enforce the IF. On the 
contrary, the presence of the random field tends to decrease the difference of the entropy of the SG and PM phases at the 
first order phase transition. As a consequence, the random field not only weakens the reentrance, but also tends to suppress 
the PM/SG first order transition corresponding to the IF by lowering the tricritical point. Although, our result is limited 
to a specific magnetic model, it could be an indication that frustration more than the disorder, would be a key ingredient to 
produce such counterintuitive type of phase transition.



\section*{Acknowledgments} 
This work has been partly supported by CNPq, CAPES, FAPERJ and FAPERGS (Brazilian agencies).

\end{document}